\documentstyle[prl,aps,twocolumn,floats]{revtex}
\begin{document}
\draft
\vskip 0.5cm
\title
{Final State of Spherical Gravitational Collapse and Likely
Source of Gamma Ray Bursts}

\author{Abhas Mitra}

\address{Theoretical Physics Division, Bhabha Atomic Research Center,\\
Mumbai-400085, India\\ E-mail: amitra@apsara.barc.ernet.in}


\maketitle

\begin{abstract}
Following our  result that for the final state of continued spherical
gravitational collapse, the gravitational mass of the fluid,
$M_f\rightarrow 0$, we show that  for a physical fluid 
the eventual value of
$2GM_f/R_f\rightarrow 1$ rather than $2GM_f/R_f <1$, indicating approach to a zero-mass black hole.
We also indicate that as the final state would be approached,
 the curvature components tend to blow up, and the proper
radial distance $l$ and the proper time $\tau \rightarrow
\infty$. This  indicates that actually the singularity is never attained
for the collapse of an isolated body.
We also identify that, the final state may correspond to the local 3-speed
$v\rightarrow c$, eventhough the circumference 
speed $U\rightarrow 0$.  However, at a finite
observation epoch,  such Eternally Collapsing Objects (ECOs) may have a
modest local speed of collapse $v \ll c$, and the lab frame speed of
collapse should practically be zero because of their extremely high
surface gravitational red-shifts.

\end{abstract}


By analyzing the GTR collapse equations for a spherically symmertic
physical fluid described by the interior comoving
metric\cite{1}:
\begin{equation}
ds^2=e^{2\phi} dt^2 -e^{2\lambda} dr^2 - R(r,t)^2
(d\vartheta^2+\sin^2\vartheta d\varphi^2)
\end{equation}
where  $R$ is the circumference
coordinate and $N(r)$ is the total number of baryons
 inclosed within the sphere of constant coordinate radius $r$,
we have shown in a recent paper\cite{2},  that
(i) no trapped surface is ever formed and
if  fluid indeed undergoes continued collapse,
the final state must have $M_f\equiv 0$ as
$R_f=0$. This result depends solely on the global properties of the field equations
manifest through the parameter: $\Gamma\equiv {dR/ dl}$,
where  $dl= e^\lambda dr$ is an element of
proper radial distance.
The local
ordinary radial velocity of the fluid in terms of comoving coordinates
$v\equiv dl/d\tau \le 1$. Here $d\tau=e^\phi dt$ is an element of proper time.
 The contraction (or expansion) rate of the invariant circumference is
$U\equiv dR/d\tau$, so that the above three functions are related by
\begin{equation}
U=\Gamma v
\end{equation}
Simultaneously, $\Gamma$
observes the following global relationship:
\begin{equation}
\Gamma^2=1+U^2 -{2G M\over R}
\end{equation}
Though, the actual collapse equations are extremely complicated and
nonlinear and defy
{\em exact} or correct solutions either analytically or numerically,
nevertheless, they are interwined globally through $\Gamma$ and $U$.
The foregoing equations can be combined in a master global equation
\begin{equation}
{\Gamma^2\over \gamma^2} =1 -{2 G M\over R}
\end{equation}
The fact $M_f=0$ means that the ``gravitational mass defect'' $\rightarrow
M_i c^2$, the original mass-energy of the fluid. And this becomes possible
when the energy liberated in the process (as measured by a distant
inertial observer $S_\infty$), $Q\rightarrow M_i c^2$. It is conceivable
that just like gravitational collapse to a neutron star (NS) stage
triggers  supernova events\cite{3}, the collapse of sufficiently massive
stellar cores triggers cosmological gamma ray bursts\cite{4}.

Our result that the assumtion of formation of ``trapped surfaces''\cite{5}
 is not
realized confirms Einstein's idea\cite{6} that Schwrzschild singularities can not
occur in practice. However, Einstein's exercise was based on the static
solutions and the theoretical formalism for GTR collapse including
pressure was developed in the sixties, much after Einstein's exercise.
The key to understand our main result is that, although $M_f=0$, the
baryon number $N$ and the baryonic mass $M_0=m N$ (as seen by $S_\infty$)
 remains fixed during the collapse process.
 And it may indeed be possible to
pack baryons indefinitely closely to achieve a state $2GM_0/R >1$ and in
fact to chase the limit $2G M_0/R\rightarrow \infty$. In Newtonian physics
the gravitational mass $M\equiv M_0$, and thus, Newtonian physics may
admit of a BH. It is small wonder then that {\em the concept of a BH
actually arose almost two hundred years ago}\cite{7}. And the idea that
the formation of a  ``trapped surface'' with $2GM/R >1$ is most natural is
thus deeply ingrained into the intuitive (Newtonian) notions which do not
distinguish between baryonic  and gravitational mass. In
contrast, in GTR,  the total mass energy $M \sim M_0 +E_g + E_{\rm
internal}+ E_{\rm dynamic}$; where the gravitational energy is negative,
and is some evolving nonlinear function of $M$, $E_g= -f(M, R)$. In the
limit of weak gravity $f \sim GM^2/R$, but as the collapse proceeds, the
grip of self-gravity becomes tighter and $M$ starts becoming reasonably
smaller   than $M_0$ or $M_i$. And if it were possible for $M$ to assume
negative values, for continued gravitational collapse, the non-linear
$E_g$ would relentlessly push $M \rightarrow -\infty$. However, there are
positive energy theorems\cite{8}, which state that the mass-energy of an {\em
isolated body} can not be negative, $M\ge 0$, and, physically, which means
that, gravitation can never be {\em repulsive}. And hence, the continued
collapse process comes to a {\em decisive end} with $M_f=0$.

 We have found that, as to the
final state, the master Eq.(4) admits only four types of solution:
\begin{equation}
1.~ \Gamma_f^2=\Gamma_i^2=U_f^2 >0,   \qquad v_f^2=1, \qquad 2GM_f/R_f=1.
\end{equation}
\begin{equation}
2.~ \Gamma_i^2>\Gamma_f^2 >0, \qquad U_f=v_f=0, \qquad 2G M_f/R_f <1
\end{equation}

\begin{equation}
3.~ \Gamma_f^2=U_f^2 =0,   \qquad v_f^2=1, \qquad 2GM_f/R_f=1.
\end{equation}
\begin{equation}
4.~  \Gamma_f^2=U_f^2 =0, \qquad  v_f=0, \qquad 2G M_f/R_f =1
\end{equation}
1. We discussed in Paper I that,
for the dust solutions, $\Gamma_f$ remains pegged to its initial value
$\Gamma_i$. Physically, this is so because, a dust solution is necessarily radiationless,
and comparable to free
fall of a test particle, for which $\Gamma=\Gamma_0= conserved~ energy~ per
unit~ rest~ mass$. And there is no question of $\Gamma$ assuming a negative
value in such a case.  

The first solution with a fixed $\Gamma= \Gamma_i=\Gamma_f$ thus
 corresponds to the $p\equiv 0$ dust solutions\cite{2}. But
we have seen that Oppenheimer-Volkov (OV)equation\cite{9} demands that for such dust
solutions $\rho\equiv 0$, and consequently, not only $M_f=0$, but also
$M_i=0$. This is possible in a strict sense only if $N\equiv 0$.

On the other hand, note that
the other  classes of solutions correspond to $U_f\rightarrow 0$. 
A physically meaningful final state must have $U_f=0$ because of the
simple the fact that the final state, by definition, must correspond to an
extremum of $R$ and therefore, we must have $U_f=dR/d\tau \equiv 0$.
However, it does not mean that the march towards the final state 
would be monotonically
decelerated for a real fluid. Depending on the unpredictable actual solutions of the
complicated non-linear coupled collapse equations and the radiation
transfer mechanism, the system may even undergo phases of acceleration and
deceleration.  In contrast,  a dust-collapse picture,  would
suggest monotonous acceleration and no emission of radiation. In such a picture,
a NS would be born on a free fall time scale
$<1$ ms. But actual microphysics intervenes and ensures that, the $\nu$
signal heralding the birth of the NS is dictated by radiation diffusion
time scale, which, in this case is $\sim 10$s.

And
since it is $U$ and not $v$ {\em which appears in the collapse equations}, in a
sense, the final state here corresponds to the static Oppenheimer-Volkov limit:
even if one would have $v\rightarrow c$!
\begin{equation}
\partial R/\partial r \rightarrow 1; \qquad R\rightarrow 0,
\end{equation}
\begin{equation}
{-g_{\rm rr}}^{1/2} =e^\lambda ={1\over \Gamma} {\partial R\over \partial r}
\rightarrow {1\over \Gamma},
\end{equation}
\begin{equation}
{d M\over dR} \rightarrow 4\pi R^2 \rho; \qquad R\rightarrow 0,
\end{equation}
and
\begin{equation}
{dp\over dR} \rightarrow -{(\rho+p)(4\pi p R^3 +M)\over \Gamma R^2}.
\end{equation}

The case (2)  corresponding to $2GM_f/R_f <1$ with $v_f =0$ indicates the 
formation of
static Ultra Compact Objects (UCOs) of  finite size $R_f$. We shall show
shortly that in case one insists for a $R_f=0$  solution in case (2), it would
imply an UCO with zero baryon number.

As far as true singular states are concerned,  the meaningful solutions
are (3) and
(4) with $\Gamma_f =0$, because, for
a physical fluid, this null value of $\Gamma_f$ implies that $M_f=0$ even
though $M_i >0$. And out of these two probable cases, the solution (3)
 is of
particular astrophysical importance. Since $v_f\rightarrow constant$
(either 0 or $c$)
for all fluid elements, it signifies complete kinetic ordering of the
fluid and, hence, eventual zero entropy. In particular, solution (4)
might correspond to a locally static $(v_f=0)$ configuration where all the
fluid particles form a single coherent perfectly degenerate system. In
this case too, the entropy of the system would be $\sim \ln(1) =0$.

In general, we see that
that $\Gamma_f \ge 0$, and $\Gamma_f$  can not assume any
negative value. This conclusion could have been directly drawn from Eq. (1)
as well. If the fluid is collapsing (expanding), on physical grounds both $U$ and $v$
must be negative (positive), implying $\Gamma \ge 0$. Also by using the general condition that
$M_f \ge 0$, it can be shown that $\Gamma_f \ge 0$.

 To some extent, the  solutions  for case 2. have already
been discussed in the literature\cite{9,10}. OV
used ideal Fermi-Dirac EOS and sought solutions with central density
$\rho_c=\infty$ and central pressure $p_c=\infty$:
\begin{equation}
\rho \rightarrow {3\over 56 \pi R^2},
\end{equation}
\begin{equation}
p\rightarrow {1\over3} \rho,
\end{equation}
\begin{equation}
M\rightarrow {3\over 14 G} R,
\end{equation}
\begin{equation}
\Gamma=constant=(4/7)^{1/2}
\end{equation}
Here it should be remembered that, as long as $R(t)\neq 0$, i.e., when one is
dealing with a non-singular configuration which may nevertheless harbor a
singularity at the center, the usual boundary condition that (i) $p=0$, at
$R=R_0$, the external surface, must be honoured. And only when a complete
singular state is reached with $R_0 =0$, the solution must be allowed to be
be truly discontinous in $p$ and $\rho$. This shows that a strictly
correct solution of the above referred OV solution could be of two types:
(1) $R_0 =\infty, ~M(R_0)=\infty$, which is an unstable solution except
for photons; (2) $R_0=R=0, ~M(R_0)=0$, and this the only stable solution
for baryons. This conclusion also follows from the solution of Misner \&
Zapolsky\cite{10} which shows that $M_{\rm core} =0, ~R_{\rm core} =0$ when
$\rho_c$ is indeed $\infty$.
And, though,
the latter OV solution is a stable one, it is devoid of any physical
content because, as was specifically discussed by OV, it has $N=0$. This
can be verified
in the following way. Note that since luminosity $L\rightarrow 0$ in the
static limit, we
must have $p_{\rm radiation} \rightarrow 0$ as $R\rightarrow 0$, so that
the total pressure $p=p_{\rm matter}$ in this limit. In the low density
limit degenerate fermions have an EOS $p \propto n^{5/3}/ m$. Since this
EOS depends on the value of $m$, it is possible that in the low density
limit, it depends on the specific nature/variety of the fermions. In
contrast, as $\rho \rightarrow \infty$, the fermi EOS becomes independent
of $m$
\begin{equation}
p =0.25 \sqrt{3/8\pi} h c n^{4/3}
\end{equation}
where $h$ is Planck's constant. In fact, it may be shown that black body photons tend to obey the $p
\propto n^{4/3}$ EOS and it is likely that all extremely degenerate bosons with a finite value
of $m$ tends to follow this EOS in the limit $\rho/m \rightarrow \infty$.
In any case for the fermions, Eqs. (12), (13), and (17) would suggest that
\begin{equation}
n=n_0 R^{-3/2}
\end{equation}
where $n_0$ is an appropriate constant. Note that for static solutions,
 by using Eqs. (10) and (18), we have
\begin{equation}
N=  \int_0^{R_0} 4\pi n e^{\lambda} R^2 dr ={4\pi
n_0}\int_0^{R_0} {R^{1/2}\over \Gamma} dR
\end{equation}
It can be easily verified that for a constant $\Gamma\neq 0$, $N=0$ as
$R_0\rightarrow 0$. Even when one does not impose the EOS (12), we have
found that the all singular solutions for the $\Gamma_f >0$ case lead to
$\Gamma=constant$ and correspondingly, $N=0$. It can be verified that,
even if we used a more general form of Eq. 8 (17) like $p \propto
n^{4+\alpha\over 3}$, for the $\Gamma_f >0$ case all singular solutions
have $N=0$ if $\alpha <1$. Therefore we discard the $2GM_f/R_f <1$ case
as physically meaningless and {\em this conclusion corroborates the spirit
of the Cosmic Censorship Conjecture} that spherical collapse can not give
rise any naked singularity\cite{11}.

With reference to the cases (3) and (4),  by inspection, we have found that the
following class of solutions are suitable :
\begin{equation}
\rho \rightarrow {1\over 8\pi G R^2}; \qquad R\rightarrow 0, \qquad
U\rightarrow 0,
\end{equation}
\begin{equation}
p\rightarrow {a_1   \over 8\pi G R^2}; \qquad a_1>0,
\end{equation}
\begin{equation}
{2 G M\over R} \rightarrow 1 -\epsilon A R^{1+2 a_2};  \qquad a_2 <0.5,
\end{equation}
where $\epsilon \ll 1$ is a tunable constant. The normalization
constant
\begin{equation}
A= R_0^{2-2 a_2},
\end{equation}
\begin{equation}
\Gamma \rightarrow \left(1-{2G M\over R}\right)^{1/2} \rightarrow \sqrt
{\epsilon A} R^{a_2 +0.5},
\end{equation}
\begin{equation}
M \rightarrow {R\over 2G}.
\end{equation}
By using Eq. (19) and (24), we find that
\begin{equation}
N={4\pi n_0\over \sqrt{\epsilon A}} \int_0^{R_0} R^{-a_2} dR
={4\pi n_0\over \sqrt{\epsilon}(1-a_2)}
\end{equation}
Now by tuning the value of $\epsilon$ one can accommodate arbitray number
of baryons in the singular state!
Recall that the for the external Vaidya metric, the dynamical collapse solution obeys the following boundary
condition at the outer boundary\cite{1,2}: $\sqrt{g_{00}}=e^\psi= \Gamma +U$
in order that $\psi \rightarrow 0$ as $R\rightarrow \infty$, i.e, to have
an asymptotically flat spacetime. Then as $R_0\rightarrow R\rightarrow 0$
and $U\rightarrow 0$, we find that
\begin{equation}
e^\psi =\sqrt{g_{00}} \rightarrow 0
\end{equation}
This shows that the {\em total surface red-shift $z\rightarrow \infty$ even
though $M\rightarrow 0$}. Similarly, by using Eq. (25),  we find that, the typical curvature component
\begin{equation}
R^{\vartheta \varphi}_{\vartheta \varphi} ={2G M\over R^3} ={1\over
R^2}\rightarrow \infty
\end{equation}
as $R\rightarrow 0$.
Also note that the proper volume of the fluid:
\begin{equation}
\Omega =\int_0^{R_0} {4 \pi R^2 \over \Gamma} dR \sim {1\over\sqrt{\epsilon}}
R_0^{1.5}\rightarrow 0
\end{equation}
since $\epsilon$ is finite.
Nevertheless, most importantly, by using Eqs. (9), (10), and (24), it
follows that, the proper radial distance to be travelled to reach this singular
state is
\begin{equation}
l \ge \int_0^{R_0} {dR\over \Gamma}\sim {1\over\sqrt{\epsilon}}
R_0^{-0.5}
\end{equation}
If $\epsilon \neq 0$, i.e., if $N$ is finite, $l \rightarrow \infty$ as $
R_0 \rightarrow 0$!  Correspondingly, the proper time required to attain
this state would also be infinite
\begin{equation}
\tau =\int {dl\over v} \ge {l\over c} \rightarrow \infty
\end{equation}
This means that time like  radial geodesics are never terminated
and
{\em the fluid can never
manage to attain this state of singularity} although it can strive to do
so continuously.  Now recall that, the most ideal condition for formation
of a BH or any singularity is the spherically symmetric collapse of a
perfect fluid not endowed with any charge, magnetic field or rotation.
If no singularity can ever be formed  under this most
ideal condition, it is almost certain that any other realistic and more
complicated case of collapse will be free of singularities. This means
that as far as dynamics of isolated bodies are concerned {\bf GTR may indeed be
free of any kind of singularities}. Even if a BH were formed its zero
surface area and would imply zero entropy content.
Thus spherical gravitational collapse does not entail any loss of
information, any violation of baryon/lepton number conservation (at the
classical non-quantum level, i.e, excluding CP violation etc. etc.) or any other
known laws of physics. The {\em fluid simply tries to radiate away its
entire mass energy and entropy in order to seek the ultimate classical
ground state}.

  Therefore, it is
probable that, subject to the presently unknown behaviour of the arbitrary
high density nuclear EOS and unknown plausible phase transitions of
nuclear matter under such conditions, GTR gravitational collapse may find
one or more quasi-stable ultracompact static configurations. Recall that
for strictly {\em cold and static equilibrium configuration} there is an
absolute upper limit on the value of the gravitational redshift, $z_{\rm
cold} <2$\cite{12}. And this would be the limiting value of $z$ for 
our static UCOs.

Nonetheless
note that, in the context of our {\em hot and dynamic solutions}, in principle,
there is no upper limit  on the probable value of $z < \infty$ (of course, if the universe has
a finite age, there would be a finite cutoff on the value of $z$). We may
remind that, in the
dynamic case $z$ would be
 a combination of
both gravitational and Doppler red-shifts.  
Permitting  likely QCD phase transitions, which might support a {\em cold
and static} ultracompact configuration with $z= z_{\rm cold}<2$, any stage
beyond it will necessarily be a dynamic stage - a march towards the
singular state described in this paper.

 This fact that $U_f\rightarrow 0$
is of great phyical significance. A distant observer is unaware of the
local 3-velocity $v=dl/d\tau$. He would practically interpret this state
with that of a {\em static} body and the apparent three velocity measured by him
$v_\infty=dR/du\rightarrow 0$, where $u$ is the  Vaidya's time or retarded time\cite{1}.

Now moving to a finite observation epoch and not necessarily to the
ultimate state, one may have the following scenario. A sufficiently
massive body, having crossed the $z <2$ limit would try to pierce through its
instantaneous Schwarzschild surface. But in order that the local (comoving)
collapse speed $v\le c$, it can not do so. Therefore, at any finite time,
the body would hover just over its instantaneous Schwarzschild
 radius ($R_g$) and
its surface gravitational redshift could be extremely high: $10$, $10^2$, $10^3$, $10^5$
or any {\em finite} number. The difference, in terms of $R$, between its
outer surface and its instantaneous Schwarzschild radius could be a
exremely small:
\begin{equation}
{R\over R_g} -1 \sim 10^{-6},~~10^{-10}
\end{equation}
or any small but {\em finite} number. In such a case, the true measure between the spatial
seperation between the outer surface of the body ($R$) and its
instantaneous Schwarzschild radius ($R_g$) is more appropriately given by
surface redshift ($z \sim 10^3, ~10^5$ or any large finite number) and the
red shift corresponding to the $R=R_g$ ($\infty$). Depending upon
the situation, i.e, whether it is a supermassive  galactic condensation or
a stellar mass ECO, and its age and other unknown details,
 the local velocity of collapse for the outer surface could be very large
$v \sim c$ or even very small, say,  $v \sim 1$ cm/s. 
The speed of collapse perceived 
by the distant observer, however, would be neglible: $V \sim dR/du \sim
\Gamma^2 ~v \sim g_{TT} ~v \approx 0$ where $g_{TT}$ is the approximate
temporal Schwarzshild metric coefficient on the surface. So, in lab frame
the object would be seen to be frozen to its instaneous Schwarzschild radius.

 And note that by POE, local physics
and kinemetics will be decided by $v$ and not by $V$. And consider an extreme case where $v \approx c$, locally and internally, the collapse may be accelerating to
the ultimate  limit $v_1\approx 1$! Now suppose
there is advection dominated accretion, with $v_2 \approx c$, onto such an (internally) Eternally
Collapsing Object (ECO). The advection dominated flow itself would cause
little accretion luminosity in case $v_1 \approx v_2$. On the other hand, if one would imagine the
ECO to be an
internally static ($v=0$) UCO, one would expect that most of the accretion
power should eventually be liberated from the ``hard surface'' of the UCO.

 And in case the observed luminosity is much lower than what is expected,
one would naturally conclude that, the flow is impinging onto an Event
Horizon rather than onto any ``hard surface''. There may be yet another reason
due to which accrretion luminosity could be much smaller than what is expected
from a ``hard surface''. If the ECO concerned is very massive, say, $10^8
~M_\odot$, its surface density could be very low, say, $10^{-4}$ g/cm$^3$.
For such a ``soft'' surface accretion flow would not produce much surface luminosity.
The flow can simply penetrate inside and increase the gravitational mass
of the ECO.

In general,  for an ECO  
with  a very large value of  $z=$, the physics of accretion
onto such a compact object may be substantially different from the present
accretion physics which  aims to study either objects with $z\ll 1$, like
WD and NS or objects with $z=\infty$, i.e., supposed finite mass BHs. It
is likely the unusual features of such high $z$ accretion process may be
interpreted as signatures for the ``detection of black holes'' in some of
the X-ray binaries and AGNs. 

Also note that,  the microphysics of
relativistic collisionless plasma is actually a poorly understood subject
and there are bound to be tacit assumptions and simplifications in all
accretion theories.
Further, because of the lack of
unambiguous observational diagnostics for the innermost region of the
central compact objects, it is really not possible at this stage to
obsevationally confirm whether the central object is a high $z$ object or a supposed BH.
But, what we can predict  with certainty is that events like
supposed ``explosion of primordial BHs''\cite{13} would never be detected.

All ECOs are expected to produce some intrinsic luminosity due to gravitational
contraction. Depending on the value of $v$ and other details, the
intrinsic radiation may lie (locally) in  X-rays or optical or other band.
However, because of the extremely large value of $z$, the radiation
observed by the distant observer would fall in the optical, radio or
microwave band. And the observed luminosity would also be smaller by the factor
$(\Gamma + U)^2 \sim z^{-2}$. In princple, however, this radiation should
be detectable. Much more importntly, the ECOs may possess magnetic fields
whose value could be either modest (in extragalactic cases) or extremely
high (in stellar mass ECOs). In contrast, the intrinsic magnetic field of
supposed BHs is zero. And ECOs might be identified as objects different
from BHs by virtue of the existence of such intrinsic magnetic fields.

And as far as the active galactic nuclei are concerned, it
is already a well known idea that their centers may contain supermassive
stars at various stages of contraction (which can very well accrete
surrounding matter) or dense regions of star
bursts\cite{14}. However, Newtonian or Post Newtonian models of
supermassive stars can not explain the sustenance of galactic nuclei for
periords longer than few years. On the other hand, we have shown here
that, actual Relativistic Configurations (Supermassive stars or stellar
mass ECOs) may appear to be almost static for distant observers for any
amount of finite duration and yet {\em keep on contracting internally} with
modest or small value of $v$.

The result that for continued collapse 
the final state must correspond to $2GM/R =1$ actually follows even if we do not use the relationship $U= \Gamma v$. As emphasized before, since the final state is an extremum of $R$, both $U= dR/dl$ and $\Gamma =dR/d \tau$ should be zero for the final state of a physical fluid (not a dust). Then it directly follows from Eq. (2) that the final state  corresponds to $2GM/R \rightarrow 1$. And since this state, by definition corresponds to $R \rightarrow 0$, its gravitational mass $M \rightarrow 0$ too. Finally, if we demand that at sufficiently small  scales quantum mechanics
must take over GTR, the final state of continued ideal spherical collapse
might be a Planck glouble of mass $M_{\rm pl} \sim 10^{-5}$g, {\em
irrespective of the initial value} of $M_i$. We can only conjecture now
whether the globule may or may not comprise wiggling elementary strings.

\end{document}